\documentclass[a4paper]{revtex4-1}
\usepackage{graphicx}
\usepackage[dvipdfm]{hyperref}
\usepackage{lmodern}
\usepackage{epsfig}

\begin{document}
\title{Multi-photon Rabi oscillations in high spin paramagnetic impurity.}

\author{S. Bertaina}
\email{sylvain.bertaina@im2np.fr}
\affiliation{IM2NP-CNRS (UMR 6242) and Aix-Marseille Universit\'{e}, Facult\'{e} des Sciences et Techniques, Avenue Escadrille Normandie Niemen - Case 142, F-13397 Marseille Cedex, France.}

\author{N. Groll}
\email{current address: Materials Science Division, Argonne National Laboratory, 9700 S. Cass Avenue, IL 60439, USA.}
\affiliation{Department of Physics and The National High Magnetic Field Laboratory, Florida State University, Tallahassee, Florida 32310, USA}

\author{L. Chen}
\affiliation{Department of Physics and The National High Magnetic Field Laboratory, Florida State University, Tallahassee, Florida 32310, USA}
\email{current address: Department of Chemistry and Chemical Biology, Cornell University, Ithaca, NY 14853-1301, USA}

\author{I. Chiorescu}
\email{ic@magnet.fsu.edu} 
\affiliation{Department of Physics and The National High Magnetic Field Laboratory, Florida State University, Tallahassee, Florida 32310, USA}


\begin{abstract}
 We report on multiple photon monochromatic quantum oscillations (Rabi oscillations) observed by pulsed EPR (Electron Paramagnetic Resonance) of Mn$^{2+}$ (S=5/2) impurities in MgO. We find that when the microwave magnetic field is similar or large than the anisotropy splitting, the Rabi oscillations have a spectrum made of many frequencies not predicted by the S=1/2 Rabi model. We show that these new frequencies come from multiple photon coherent manipulation of the multi-level spin impurity. We develop a model based on the crystal field theory and the rotating frame approximation, describing the observed phenomenon with a very good agreement. 
\end{abstract}

\maketitle

\section{Introduction}

Quantum coherence of electron spin systems probed by pulsed EPR attracts a great interest due to its potential applications in quantum computation. Spin systems benefit from relatively large coherence and relaxation times, which makes them suitable as quantum bit implementations or as a new type of quantum random access memory \cite{Blencowe2010, Chiorescu2010, Kubo2010, Schuster2010}. Recent studies show that the dilution of spins in diamagnetic hosts significantly reduces spin-spin dipolar interactions and allows spin quantum manipulations. Such situations are well exemplified by systems like N atoms in C$_{60}$\cite{Morley2007}, NV centers in diamonds \cite{Hanson2008,Dutt2007} , Er$^{3+}$\cite{Bertaina2007,Bertaina2009a}, Yb$^{3+}$\cite{Rakhmatullin2009} and Cr$^{5+}$\cite{Nellutla2007} ions and molecular magnets\cite{Bertaina2008,Takahashi2009}.

The majority of studies on electron spin qubits uses S=1/2 isotropic systems, which prevents strong static effects due to the crystal field or unwanted non linear effects. However, coherent manipulation of a multilevel system is needed as the first step towards the implementation of the Grover algorithm\cite{Grover1997,Leuenberger2001,Leuenberger2002}. 

Although several quantum continuous wave (CW) EPR spectroscopy studies have been reported  years ago\cite{Sorokin1958}, coherent multiple photon EPR ( Rabi oscillations)  have not been clearly  demonstrated until recently \cite{Bertaina2009}.

In order to study quantum effects in multi level electronic spin systems, we chose Mn$^{2+}$ (S=5/2) ion implanted inside the diamagnetic host MgO. The cubic symmetry of MgO induces a small crystal field anisotropy on Mn$^{2+}$ and, as we will see in the following sections, the microwave magnetic energy can be brought to the same order of magnitude, an aspect shown to be instrumental in generating multi-photon Rabi oscillations.   

\section{Coherent dynamic of MgO:Mn$^{2+}$}\label{sec:spinH}

\subsection{Spin Hamiltonian}
 
The spins $S$=5/2 of the Mn$^{2+}$ ions are diluted in a MgO non-magnetic matrix of cubic symmetry $F_{m\bar{3}m}$ (lattice constant 4.216 {\AA}). The Mn$^{2+}$ ions are located in substitutional positions of Mg$^{2+}$ ions. The high degree of symmetry ensures that the spins are seeing an almost isotropic crystalline environment and thus the fourth order magnetic anisotropy can be made much smaller than the Zeeman splittings. Interactions between Mn$^{2+}$ ions and symmetry deformations are neglected. The spin Hamiltonian at resonance is given by \cite{Low1957,Smith1968,Bertaina2009}:

\begin{eqnarray}\label{eq:1}
    H&=&a/6\left[ {S_x^4+S_y^4+S_z^4-S(S+1)(3S^2-1)/5} \right]\\
&&+\gamma\vec{H}_0\cdot\vec{S}-A\vec{S}\cdot\vec{I}+\gamma\vec{h}_{mw}\cdot\vec{S}
\cos(2\pi ft) \nonumber
\end{eqnarray}
where $\gamma=g\mu_B/h$ is the gyromagnetic ratio ($g=2.0014$ the $g$-factor, $\mu_B-$ Bohr's magneton and $h-$ Planck's constant), $a =$ 55.7~MHz is the anisotropy constant, $A = 244$~MHz is the hyperfine constant of $^{55}$Mn ($I = 5/2$), $h_{mw}$ and $f$ represent the microwave (MW) amplitude and frequency respectively, and $\vec{H}_0$ is the static field ($\vec{H}_0\perp \vec{h}_{mw}$). In our experiments, the static field ensures a Zeeman splitting of $\gamma H_0\approx f\sim$9~Ghz, much stronger than all the other interactions of Eq.(\ref{eq:1}). This implies that (i) $\vec{H}_0$'s direction can be approximated as the quantization axis and (ii) coherent MW driving is confined between levels of same nuclear spin projection $m_I$.\\

\subsection{Experimental procedure}

\begin{figure}[th]
\centering
\includegraphics[width=0.8\columnwidth]{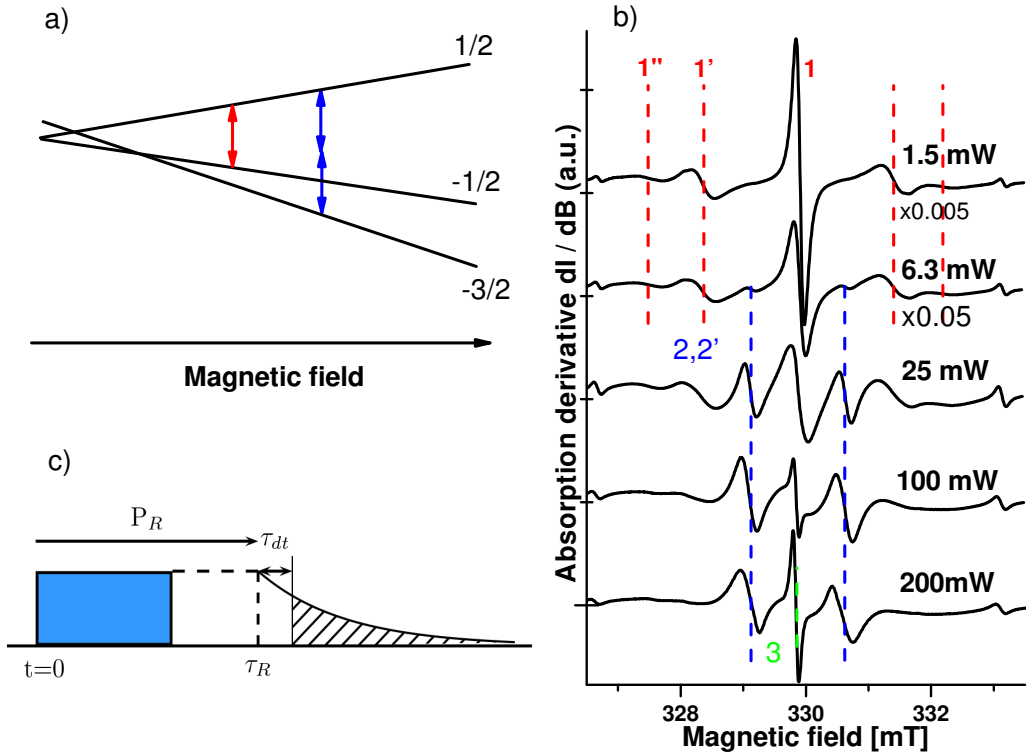}
\caption{Multiple photon resonances in CW configuration. a) Schematic representation of the Zeeman levels (only 3 levels are shown) with increasing magnetic field. The red arrow is the conventional 1-photon transition. The blue arrows represent the 2-photon transition between $|-3/2\rangle \rightarrow |1/2\rangle$. b) CW spectra of MgO:Mn$^{2+}$. At low MW power, only the 1-photon transitions are induced (red dashed lines) but when the MW power is increased, multi-photon transitions appear. c) The pulse sequence used in the multi-photon Rabi oscillation measurements, starting with a strong excitation pulse P$_R$ which induces the multi-photon nutation. Right after the P$_R$ pulse, the FID signal gives the transverse magnetization state between -1/2 and 1/2.}
\label{fig:mn_cw}   
\end{figure}

Rabi oscillations measurements have been performed in a Bruker Elexsys 680 pulse EPR spectrometer working at about $f=9.6$~GHz ($X$-band). All data have been recorded at room temperature.
Multiple photon resonance in continuous wave (CW) non coherent spectroscopy is explained in Fig. \ref{fig:mn_cw}. In pulsed EPR both the static field and MW frequency are fixed. In our experiments the static magnetic field has been chosen to be in the middle of the transition line -1/2 $\rightarrow$ 1/2. A microwave pulse $P_R$ start at $t=0$ and coherently drive the magnetization. At the end of the pulse ($t=t_P$) the magnetization must be recorded in order to be compared with the present theory. Because of the dead time of the spectrometer ($\tau_{dt}\sim$ 80ns), it is impossible to directly measure the magnetization right after the pulse $P_R$. However if the line is sharp enough we can simply record the free induction decay (FID) emitted by the system when the microwave field is shut down. This method gives the value of the density of population at the end of the pulse (Fig. \ref{fig:mn_cw}).\\

\section{Results and discussions.}
\begin{figure}[tbh]
\centering
\includegraphics[bb=0 0 186 106,clip, width=0.8\columnwidth]{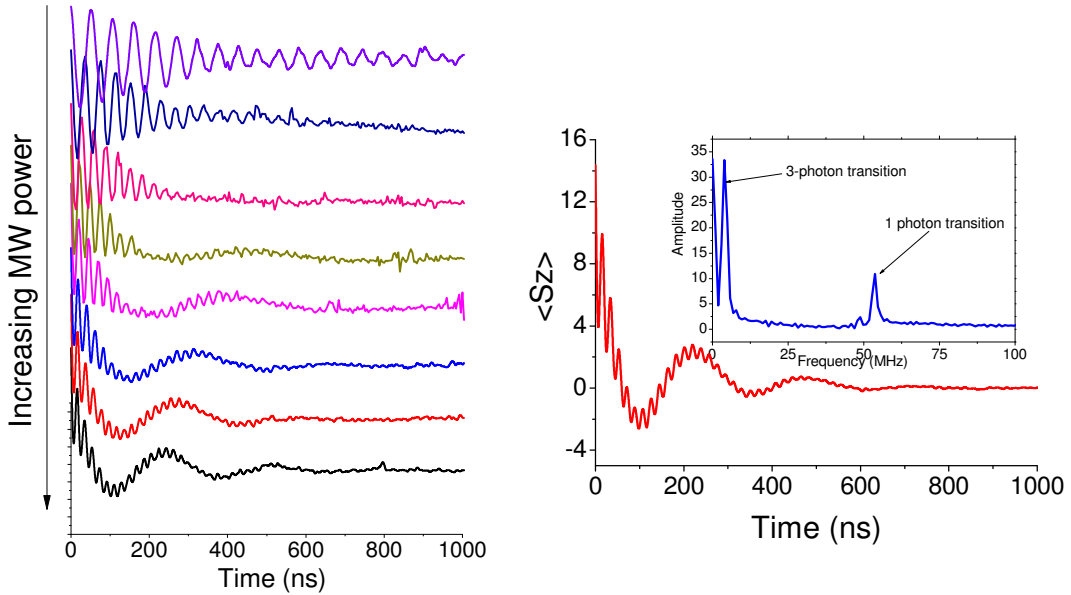}
\caption{Rabi oscillations of the Mn$^{2+}$ ions in MgO when the microwave power is increased, for $\vec{H}_0||[111]$. The FFT clearly shows 2 Rabi frequencies: the higher one is induced by a 1-photon process whereas the lower frequency is due to a 3-photon process.}
\label{fig:rabi1}   
\end{figure}

Examples of Rabi oscillations are given in Fig. \ref{fig:rabi1}. At low MW power only the 1-photon Rabi oscillation is observed. When the MW power is increased a second kind of oscillation superimposes with the one-photon process. The Fourier transform of the Rabi oscillation clearly show 2 well resolved frequencies.    

In order to fully describe the experimental data we developed a model based on the Hamiltonian (\ref{eq:1}). Since the static field is much bigger than all the other interactions of (\ref{eq:1}), the quantization axis is defined by $\vec{H}_0$. Once the static part of (\ref{eq:1}) is diagonalized, the full Hamiltonian (including the dynamical part) is transformed using the unitary transformation $U(t)=\exp(-i2\pi f \hat{S}_z t)$. Neglecting the counter rotation of the microwave radiation (rotating wave approximation) we get the Hamiltonian in the rotating frame :

\begin{equation}\label{eq:rwa}
    \mathcal{H}_{RWA}=\frac{g\mu_B}{2h} h_{mw}m \hat{S}_x+\hat{E}-f\hat{S}_z
\end{equation}  

The dynamics of Hamiltonian~(\ref{eq:rwa}) is described by the time evolution of the 6x6 density matrix $\rho(t)$:
\begin{equation}\label{eq:6}
    i\frac{d\rho}{dt}=[\mathcal{H}_{RWA}, \rho].
\end{equation}
In the rotating frame (Dirac or interaction picture), the solution is $\rho(t)=U_p(t)\rho_0 U^\dag_p(t)$ with $\rho_0$ the matrix density at thermal equilibrium :
\begin{equation}\label{eq:7}
    \rho_0=\frac{\exp(-\hat{E}/kT)}{Tr (\exp(-\hat{E}/kT))}
\end{equation}  
and $U_p(t)$ is the propagator operator: $U_p(t)=\exp(-i2\pi \mathcal{H}_{RWA} t )$.

When the Rabi pulse $P_R$ is applied, the spin population is coherently manipulated. At the end of the sequence a weak $\pi/2$ pulse will selectively probe the difference of population between states -1/2 and +1/2. The signal out of the spectrometer is $S(t)=\sigma_{-1/2}(t)-\sigma_{1/2}(t)$.

\begin{figure}[tbh]
\centering
\includegraphics[bb=0 0 140 102,clip, width=0.49\columnwidth]{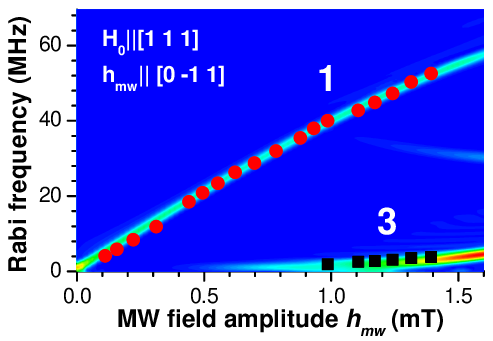}
\includegraphics[bb=0 0 140 128,clip, width=0.49\columnwidth]{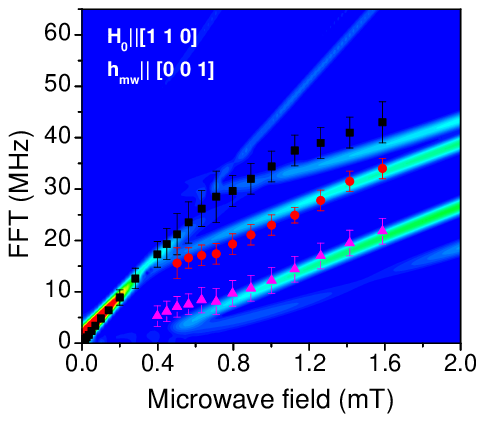}
\caption{MW field dependence of the Rabi oscillations in MgO:Mn$^{2+}$ for 2 orientations between the crystal and the static field $\vec{H}_0$. For both figures, the points are FFT peaks of the experimentally recorded Rabi oscillations. The contourplots have been computed numerically using the presented model.  }
\label{fig:rabi2}   
\end{figure}

Results of the presented model are presented in figure \ref{fig:rabi2}. The MW field dependence of the Rabi frequency of the experimental oscillations are in very good agreement with the presented model. Note that no fit parameter has been used and all parameters have been take from previous independent studies \cite{Low1957,Smith1968}. For $\vec{H}_0||[111]$ the frequency distribution is quite simple showing 2 kinds of Rabi oscillations: 1-photon and 3-photon. For  $\vec{H}_0||[110]$ the frequency distribution is more complicated but still well describes by the model. In this case, there are 1-photon (Black squares) and 2 kinds of 3-photon process (red circles and purple triangle), coupling resonantly different levels of the 6-level spin system.

We note that static field angular dependence studies \cite{Bertaina2009,Bertaina2011}, not presented here, also shows also a very good agreement between this model and the experimental data.

\section{Conclusion} 

We report on single and multi-photon coherent rotations in a $S=5/2$ spin system featuring a cubic anisotropy. We introduce a model, using only the spin Hamiltonian, that describes the experimental data with a very good agreement. These result are an important step toward the implementation of the Grover algorithm which need the coherent manipulation of a multilevel system.
 
\section*{Acknowledgments}
This work was supported by City of Marseille, NSF Cooperative Agreement Grant No. DMR-0654118, NSF grants No. DMR-0645408, Aix-Marseille University (BQR grant).

\section*{References}


\end{document}